\documentclass[doublecol,figures]{epl2}
\usepackage{amsmath,amssymb}

\title{Adaptive model for recommendation of news}
\author{Mat\'u\v s Medo\inst{1}\thanks{E-mail:
\email{matus.medo@unifr.ch}} \and Yi-Cheng Zhang\inst{1} \and
Tao Zhou\inst{1,2}}
\shortauthor{Mat\'u\v s Medo \etal}

\institute{
  \inst{1} Physics Department, University of Fribourg,
           CH-1700~Fribourg, Switzerland\\
  \inst{2} Department of Modern Physics, University of Science and
           Technology of China, Hefei 230026, P. R. China}


\abstract{Most news recommender systems try to identify users'
interests and news' attributes and use them to obtain
recommendations. Here we propose an adaptive model which combines
similarities in users' rating patterns with epidemic-like spreading
of news on an evolving network. We study the model by computer
agent-based simulations, measure its performance and discuss its
robustness against bias and malicious behavior. Subject to the
approval fraction of news recommended, the proposed model
outperforms the widely adopted recommendation of news according
to their absolute or relative popularity. This model provides a
general social mechanism for recommender systems and may find its
applications also in other types of recommendation.}

\begin{document}
\maketitle

\section{Introduction}
People were always hungry for information. To satisfy their
needs, many information sources have been created and now they
are competing for our attention~\cite{Gold97,Huber}. News
distribution on the Internet is also still fashioned the old,
centralized way. Even the new services like digg.com,
reddit.com, wikio.com and others, where the traditional news
distribution paradigm is challenged by assuming that it is the
readers who contribute and judge the news, have a serious
drawback: every reader sees the same front page. As a result,
only news items of very general interest can become popular.
Niche items, \emph{i.e.} those that target a particular interest
or locality, do not have much chance to reach their audience.

An~alternative approach is to deliver ``the right news to the
right readers'' as provided by systems for adaptive news
access~\cite{BiPa07}. These systems accommodate the interests of
their users and provide a~personalized set of interesting news
for each individual. They reflect their readers' actions by
either \emph{news aggregation} (where each user can choose
preferred sources and topics), \emph{adaptive news navigation}
(this is achieved mainly by creating lists of most popular
news---a technique which is adapted by most newspaper websites
but can be implemented also in a more sophisticated way as
recently suggested in~\cite{HuWu07}), \emph{contextual news
access} (providing news according to the currently viewed
information), or by \emph{content personalization} on the basis
of past user's preferences. The last option mentioned is a
specific application of recommender systems---a widely-applied
tool for information filtering~\cite{HKTR04,AdoTu05}.

Various systems for personalized news recommendation were
proposed in past. Possibly the simplest approach, known also as
``collaborative filtering'', is based on using the correlations
between users' ratings~\cite{RISBR94}. Often used is learning
the keywords of interest for each individual user and
recommending the news that contain them~\cite{BPCh00}.
Similarly, when both news and readers' interests can be
described by a~couple of categories, recommendations can be
obtained by matching news's attributes with user's
preferences~\cite{Mussi03,CBC08}. Most news recommender systems
are constructed in this way, only that the handful of categories
is replaced by a more general user's reading profile which is
inferred from the user's feedback on previously read
news~\cite{LLK03,PCBC07,ABGHS07}. In some cases, separate models
addressing user's short-term and long-term interests are used
and the final recommendation is obtained as a~mix of the two
results~\cite{BiPa00}. Explicit user ratings of news can be
replaced by implicit ratings (for example, the mere access of a
news may be interpreted as the user's satisfaction) or by
ratings inferred from reading times (when ``short'' and ``long''
reading times are interpreted as user's dissatisfaction or
satisfaction respectively)~\cite{LLK03}. For an overview of this
rapidly developing field see~\cite{BiPa07,CBC08}.

The news recommender model which we propose and study in this
paper is different from the systems described above. While
preserving the user-posting-news feature which is often used by
popular websites, we aim at personalized news recommendation by
observing readers' past reading patterns, identifying their
taste mates and constructing a directed local neighborhood
network. In our model, users read news and either ``approve'' or
``disapprove'' them. When a news is approved, it spreads in the
neighborhood network to the next prospective readers. This
process is similar to an epidemic spreading in a social
network~\cite{PSV01,ZFW06} or to rumor spreading in
a~society~\cite{MNP04,CFL07}. Simultaneously with the spreading
of news, the network of contacts gradually evolves to best
capture users' similarities.

To summarize, with the reading community acting as a collective
social filter, we aim to navigate news items to their intended
readership. The model's reliance on connecting the users with
similar reading interests is motivated by the basic paradigm of
recommender systems: you get recommended what your taste-mates
already liked~\cite{AdoTu05}. However, recommendation of news
has an important flavor which is missing in most other
applications of recommender systems: novelty is of crucial
importance there. In our case, the challenge of fast news
delivery is addressed by the exponentially-fast spreading of
good news (which is a direct consequence of the spreading
mechanism) while the importance of novelty is reflected by a
later introduced continual time decay of the recommendation
scores.

\section{Description of the model}
Here we describe the adaptive news recommendation model,
assuming no other information than ratings of news by users.

\subsection{Notation}
In this paper, $U$ is the total number of users, $S$ is the
number of trusted sources (authorities) assigned to each user,
and $s_{ij}$ is the estimated similarity of reading tastes of
users $i$ and $j$. We use Latin letters to label the users and
Greek letters to label the news. Evaluation of news $\alpha$  by
user $i$, $e_{i\alpha}$, is either $+1$ (liked/approved), $-1$
(disliked/disapproved) or $0$ (not evaluated yet). The
recommendation score of news $\alpha$ for user $i$ is
$R_{i\alpha}$.

\subsection{Estimation of user similarity}
User similarity is estimated from users' assessments of the news.
When users $i$ and $j$ match in evaluations of $m$ news and
mismatch in evaluations of $M$ news, the overall probability of
agreement can be estimated as $m/(m+M)$ and this number can be
used as a measure of similarity of these users. However, such an
estimate is prone to statistical fluctuations: it is the user
pairs with a small number of commonly evaluated news $m+M$ that
are likely to achieve ``perfect'' similarity $1$. Since in
sampling of $n$ trials, the typical relative fluctuation is of
the order of $1/\sqrt{n}$, we estimate the user similarity as
\begin{equation}
\label{similarity}
s_{ij}=\frac{m}{m+M}\bigg(1-\frac{\theta}{\sqrt{m+M}}\bigg)
\end{equation}
where $\theta$ is a factor determining how strongly we penalize
user pairs with few commonly evaluated news. The value $\theta=1$
yielded optimal results in our tests and we use it in all
simulations presented in this paper. When $m+M=0$ (no overlap),
we set $s_{ij}=\varepsilon$ where $\varepsilon$ is a~small
positive number: this reflects that even when we know no users'
evaluations, there is some base similarity of their interests.

\begin{figure}
\centering
\includegraphics[scale=0.9]{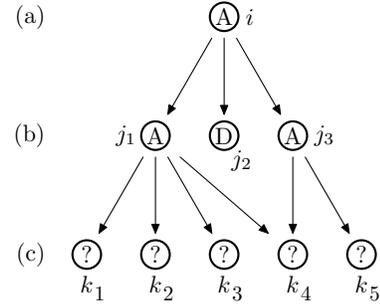}
\caption{Illustration of the news propagation. (a) User $i$ added
a~new news, which is automatically considered as approved (A) and
sent to users $j_1,j_2,j_3$ who are $i$'s followers. (b) While
user $j_2$ dislikes (D) the news, users $j_1$ and $j_3$ approve
it and pass it further to their followers $k_1,\dots,k_5$ who
haven't evaluated the news yet (which is denoted with question
marks). (c) User $k_4$ receives the news from the authorities
$j_1$ and $j_3$, yielding the news's recommendation score
$s_{j_1k_4}+s_{j_3k_4}$. At the same time, user $k_5$ receives
the news only from the authority $j_3$ and hence for this user,
the recommendation score is only $s_{j_3k_5}$.}
\label{fig:illustration}
\end{figure}

\subsection{Propagation of news}
One can use all currently available user evaluations to estimate
similarities $s_{ij}$ for all user pairs. Since the memory
needed to store the result grows quadratically with the number
of users, this is not a scalable approach to the problem. To
decrease the memory demands, we keep only $S$ strongest links
for each user. Those $S$ users who are most similar to a given
user $i$ we refer to as authorities of $i$ and, conversely,
those who have user $i$ as an authority we refer to as followers
of $i$. Notice that while the number of authorities for each user
is fixed, a highly valued user may have a large number of
followers. Lacking any prior information, we assume random
initial assignment of authorities. As the system gathers more
evaluations, at regular time intervals it evaluates the data
and selects the best authorities for each user.

The directed network of authorities and followers described
above serves as a basis for news propagation in our model. After
news $\alpha$ is introduced to the system by user $i$, its
initial recommendation score is zero for all users:
$R_{i\alpha}=0$. In addition, the news is ``passed'' to all
followers of $i$. For each such user $j$, the recommendation
score increases by $s_{ij}$ (\emph{i.e.}, the higher the
similarity with the news's originator, the stronger the
recommendation). When news $\alpha$ is approved by user $j$, it
is passed further to all followers of $j$ and for each of those
users, the recommendation score is increased by their similarity
with $j$. That means, when user $j$ approves news $\alpha$,
recommendation scores of this news are updated as
\begin{equation}
\label{score_improvement}
R_{k\alpha}'=R_{k\alpha}+s_{kj}
\end{equation}
where $k$ is a~follower of $j$. For user $i$, the available
unread news are sorted according to $R_{i\alpha}$ (high scores at
the top). As it is illustrated in Fig.~\ref{fig:illustration},
when a user receives the same news from multiple authorities, the
news's recommendation score increases multiple times and hence
the news is more likely to get to the top of the user's
recommendation list and be eventually read.

In effect, the above algorithm implies that news spread in
a~directed network of users. Since similarities $s_{kj}$ are
positive, recommendation scores updated according to
Eq.~(\ref{score_improvement}) can only grow with time which
gives unjustified advantage to old news. We shall introduce
a~time decay of the scores in the following section.

\subsection{Updating the assignment of authorities}
Authorities of user $i$ should be always those $S$ users who
have the highest rating similarity with $i$. While this requires
continual updating of authorities, as the optimal assignment is
approached, the updating can be done less frequently. For
simplicity, we update the assignment of authorities every ten
time steps in all numerical simulations.

\section{Numerical validation of the model}
We devise a~simple agent-based approach to test and optimize the
proposed model (for an introduction to agent-based modeling
see~\cite{LaKe97}). It's not our goal to provide a~perfect model
of readers' behavior. Instead, we aim to make plausible
assumptions allowing us to study the model under various
circumstances.

\subsection{Agent-based model}
To model user's judgment of read news we use a vector model
where tastes of user $i$ are represented by the $D$-dimensional
taste vector $\boldsymbol{t}_i=(t_{i,1},\dots,t_{i,D})$ and
attributes of news $\alpha$ are represented by the
$D$-dimensional attribute vector
$\boldsymbol{a}_{\alpha}=(a_{\alpha,1},\dots,a_{\alpha,D})$. We
use $D=16$ and set the taste vectors such that each user has
preference for $D_1=6$ of $16$ available tastes (hence, each
taste vector has six elements equal to one and the remaining ten
elements equal to zero). There are $\binom{D}{D_1}=8\,008$ such
vectors and hence there are $8\,008$ users in our system who all
have mutually different taste vectors. Satisfaction of user $i$
with news $\alpha$ is assumed in the form
\begin{equation}
\label{overlap}
\Omega(i,\alpha)=Q_{\alpha}(\boldsymbol{t}_i,\boldsymbol{a}_{\alpha})
\end{equation}
where the scalar product $(\boldsymbol{t}_i,\boldsymbol{a}_{\alpha})$
represents the overlap of $i$'s tastes and $\alpha$'s attributes
and the multiplier $Q_{\alpha}$ represents the overall quality
of news $\alpha$ (similar vector models are often used in
semantic approaches to recommendation~\cite{CBC08}). When a news
is introduced to the system, its attributes are set identical
with the tastes of its originator and $Q_{\alpha}$ is drawn from
the uniform distribution $\mathcal{U}(0.5,1.5)$. We assume that
user $i$ approves news $\alpha$ only when
$\Omega(i,\alpha)\geq\Delta$; the news is disapproved otherwise.

Simulation time advances in steps. We assume that in each time
step, a given user is active with the probability $p_A$. Each
active user reads top $R$ news from the recommendation list
(this is motivated by the study showing that users mostly visit
pages that appear at the top of search-engine
results~\cite{ChoRoy04}) and with the probability $p_S$ submits
a~new news.

\subsection{Performance measures}
The ratio of news' approvals to all assessments is an obvious
measure of the system's performance. This number, which we refer
to as \emph{approval fraction}, tells us how often are users
satisfied with the news they get recommended.

In the computer simulation, we have the luxury of knowing users'
taste vectors and hence we can compute the number of differences
between the taste vector of a user and the taste vectors of the
user's authorities. By averaging over all users, we obtain the
average number of differences. Obviously, the less are the
differences, the better is the assignment of authorities. Since
we assume that all taste vectors are mutually different, the
smallest number of differences is two and hence we introduce
\emph{excess differences} which is the average number of
differences minus two and the optimal value of this quantity is
zero.\footnote{When the number of authorities $S$ is large (in
our case, when $S>(D-D_1)D_1$), it's impossible to reach zero
excess differences.}

\begin{figure}
\centering
\includegraphics[scale=0.25]{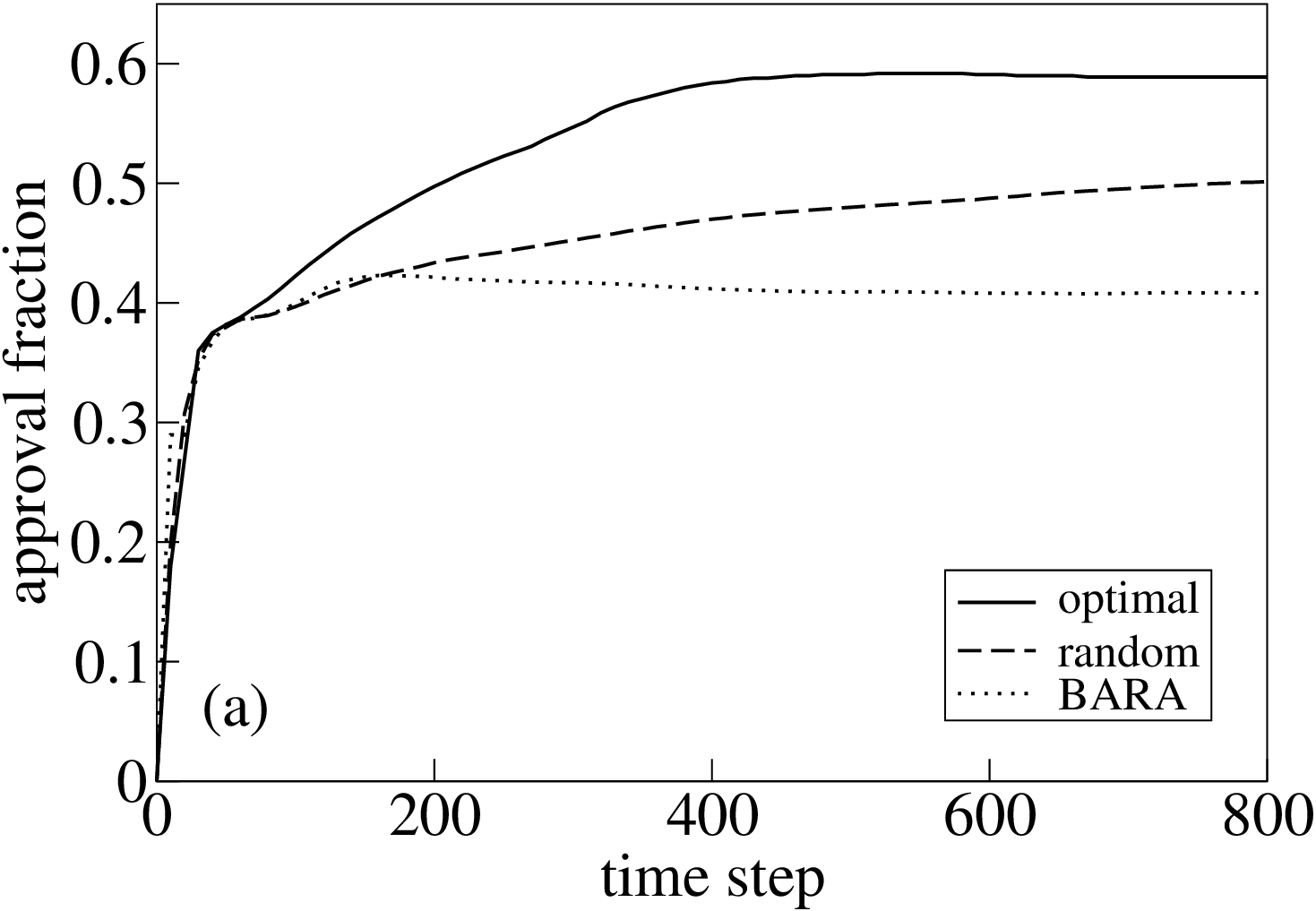}\\[8pt]
\includegraphics[scale=0.25]{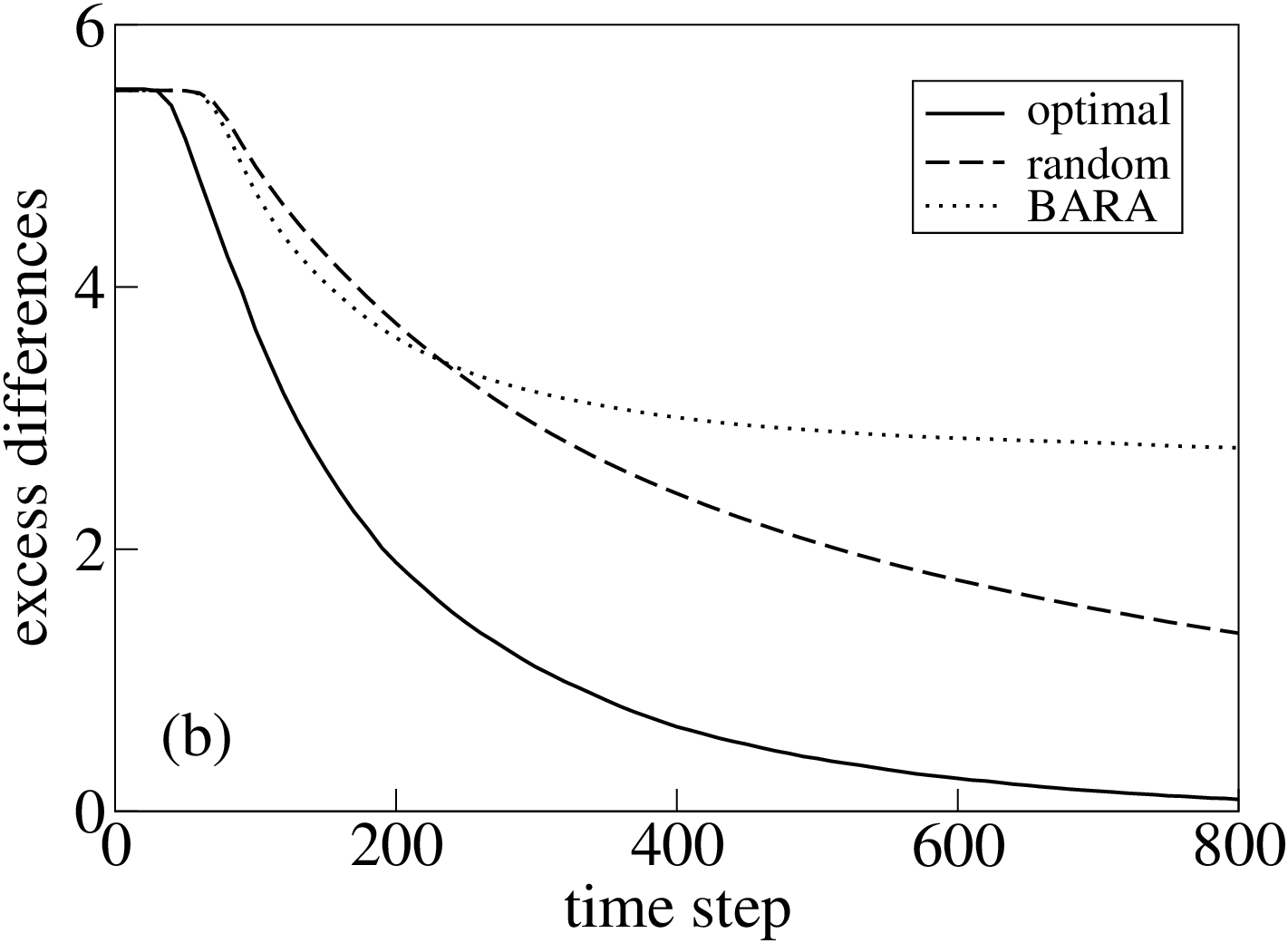}
\caption{Comparison of various rewiring procedures: approval
fraction (a) and excess differences (b) as a function of time for
optimal, random, and BARA updating of authorities (results were
averaged over ten independent realizations).}
\label{fig:var_rewirings}
\end{figure}

\subsection{Results}
First we study if the system is able to evolve from the initial
random assignment of authorities into a more appropriate state.
Instead to the described updating of authorities, one can think
of a simple ``replace the worst'' scheme: in each step, every
user confronts the least-similar authority with a~randomly
selected user $k$. When the authority's similarity with the user
is lower than $k$'s similarity (and $k$ is not the user's
authority yet), the replacement is made. Such random sampling is
obviously less computationally demanding than the original
optimal approach which, on the other hand, makes use of all the
information available at the moment. A compromise of the two
approaches is to replace $i$'s least-similar authority with one
of the users who are authorities for $i$'s most-similar
authority (hence the name ``best authority's random authority'',
BARA).

We compare the three updating rules for $S=10$ (ten authorities
per user), $p_A=0.02$ (\emph{i.e.}, on average, a user is active
every 50 steps), $R=3$ (active user reads three top news from
the recommendation list), $p_S=0.01$ (on average, one of hundred
active users submits a news), $\Delta=3$, and
$\varepsilon=0.001$. As can be seen in
Fig.~\ref{fig:var_rewirings}, the optimal choice of authorities
yields higher approval fractions and lower excess differences
than the other two methods. The worst performing is the BARA
updating---while it initially converges slightly faster than the
random sampling, it reaches only a~strongly sub-optimal
assignment of authorities. The initial plateau of the excess
differences is due to the little information available to the
system at the beginning of the simulation. The initial value of
excess differences in Fig.~\ref{fig:var_rewirings}b, $5.5$,
corresponds to the random initial assignment of
authorities.\footnote{This number depends on the parameters
chosen---denoting the number of ones in each of the
$D$-dimensional taste vectors as $D_1$, the average number of
differences can be computed as $\overline{d}=2\sum_{d=1}^{D_1}
d\binom{D_1}{d}\binom{D-D_1}{d}/\big[\binom{D}{D_1}-1\big]$.}

\begin{figure}
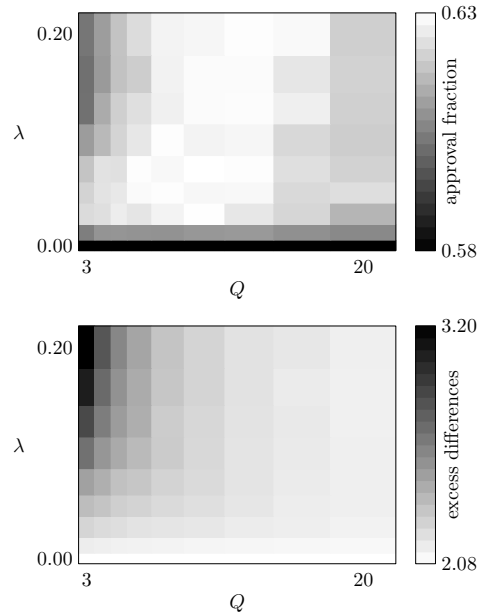

\centering
\includegraphics[scale=0.75]{density_plot-3}
\includegraphics[scale=0.75]{density_plot-7}
\caption{Effects of the time decay on the system's performance
at time step 800, when the system is almost equilibrated
(results were averaged over ten independent realizations).}
\label{fig:damping}
\end{figure}

An important flavor is still missing in the proposed model: a
time decay of news' recommendation scores. With no decay
present, recommendation scores never decrease and a news is
removed from a user's reading list only as it eventually gets
read. In addition, with many old news queued, it's hard for good
fresh news to get to the top of a recommendation list and catch
the user's attention. A simple solution for all these problems
is achieved by incremental decreasing of recommendation scores
with time. We implement the time decay in the following way: in
each time step, when a user has more than $Q$ queued news, we
decrease their recommendation scores by a small value $\lambda$
and news with $R_{i,\alpha}\leq0$ are removed from the list. As
shown in Fig.~\ref{fig:damping}, an appropriately set time decay
significantly increases the number of excess differences and
enhances the approval fraction.

\begin{figure}
\centering
\includegraphics[scale=0.25]{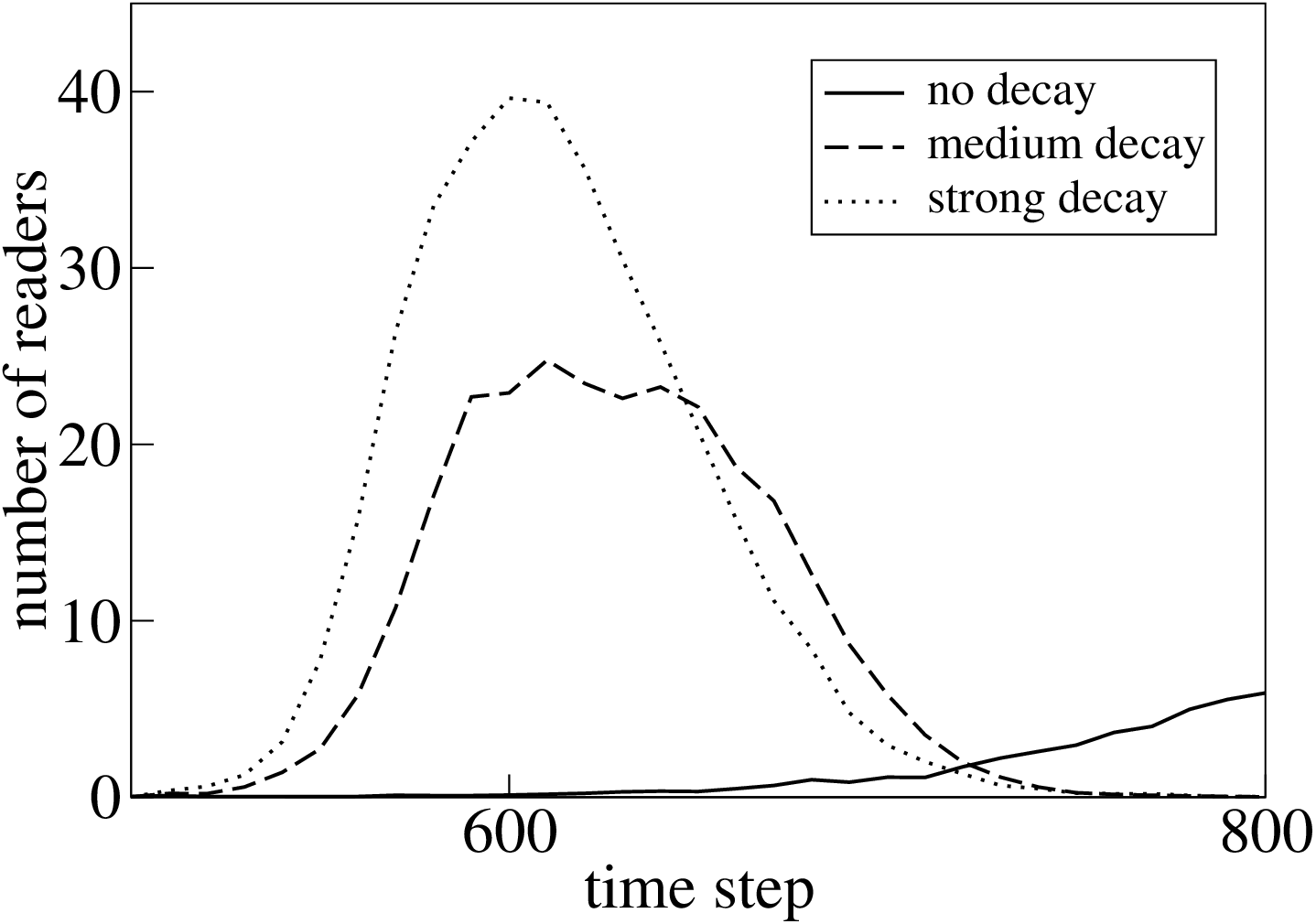}
\caption{Evolution of the number of readers per time step for
ten high quality news introduced shortly after $T=500$: no decay
($\lambda=0$), medium decay ($Q=10$, $\lambda=0.1$), strong
decay ($Q=10$, $\lambda=4.0$).}
\label{fig:time_evol}
\end{figure}

Apart from the moderate improvement of both performance measures,
the decay of recommendation scores is crucial in promotion of fresh
news. To illustrate this effect we did simulations where first ten
news introduced after time step 500 (when the system is almost
equilibrated) had particularly high qualities. We used this
setting to examine how the average attention paid to those
superior news evolves with time. As can be seen in
Fig.~\ref{fig:time_evol}, without decay, good news stay queued
for exceedingly long time before they reach their audience
(solid line). On the other hand, when the decay is too strong,
even good news may be eliminated prematurely (dotted line). As a
compromise between promotion of fresh news and two performance
measures (approval fraction and excess differences), we use
$Q=10$ and $\lambda=0.1$ in all following simulations.

\begin{figure}
\centering
\includegraphics[scale=0.25]{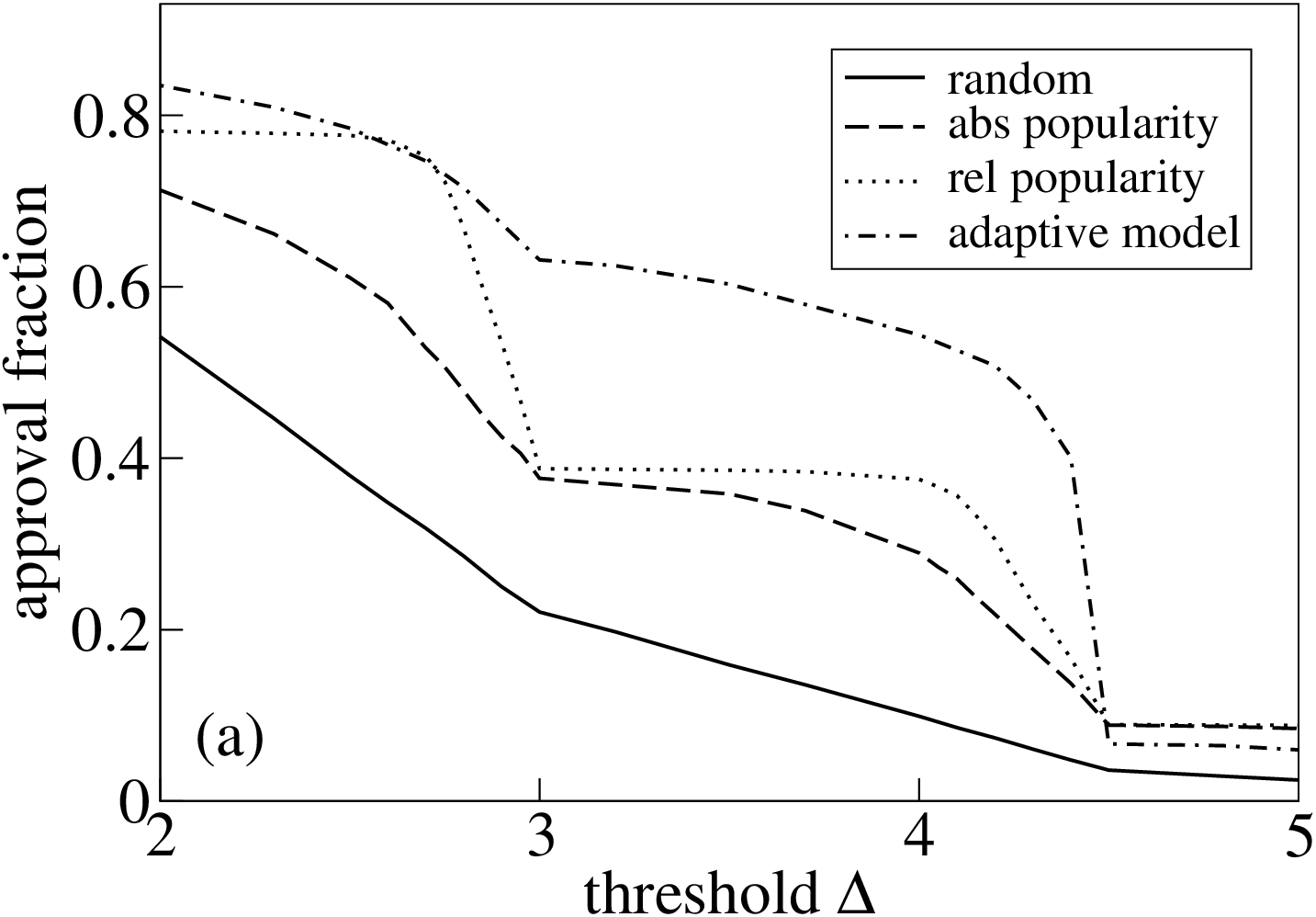}\\[8pt]
\includegraphics[scale=0.25]{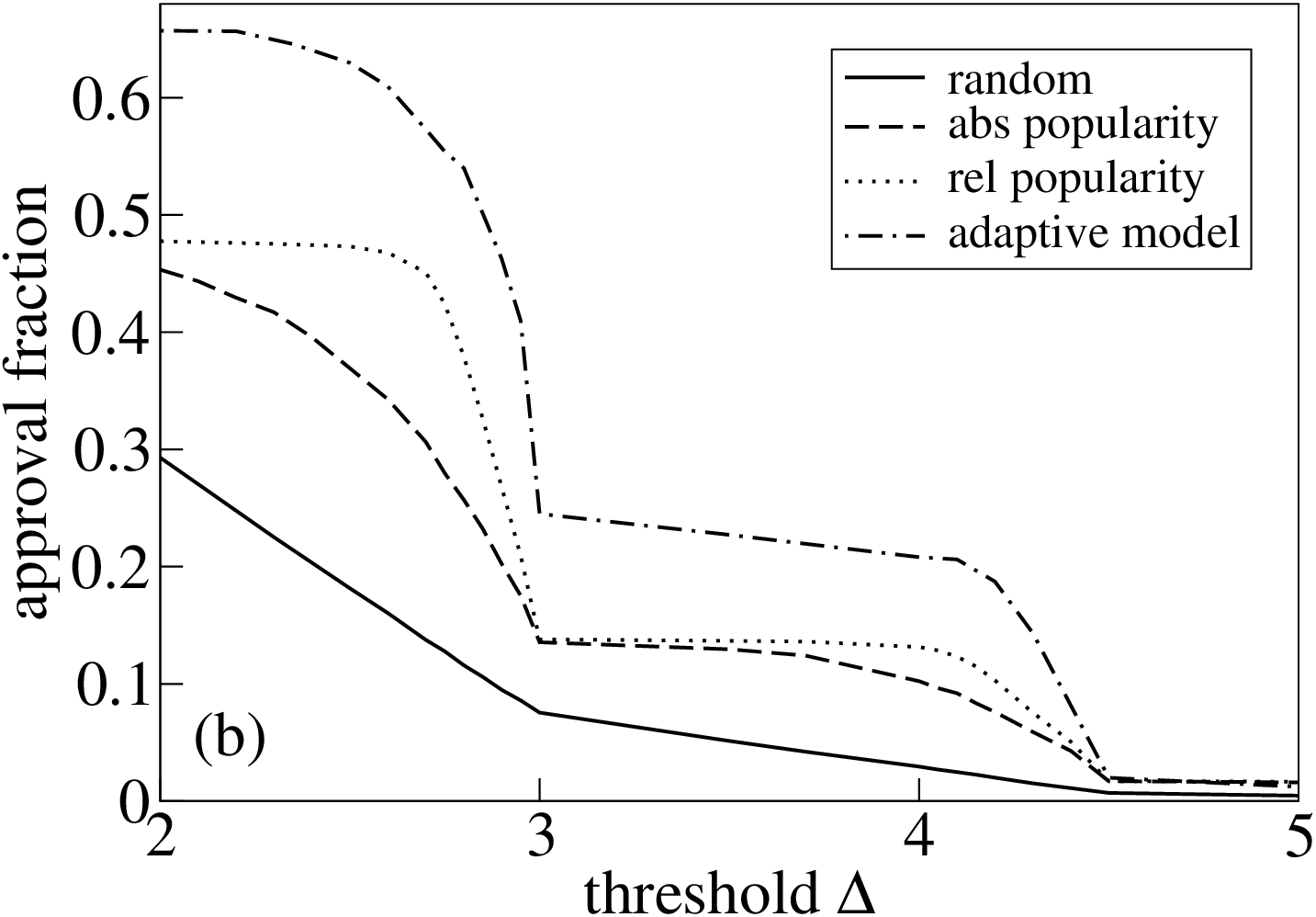}
\caption{Comparison of three simple filtering mechanisms with
the proposed adaptive model. On the horizontal axis we have the
evaluation threshold $\Delta$ which characterizes how demanding
the users are. (a) The original setting with $6$ active tastes
out of $16$. (b) The total number of tastes is $24$, only $6$ of
them are ``active''.}
\label{fig:comparison}
\end{figure}

Having seen that the proposed system is able to improve the
assignment of authorities and thus filter the news, a natural
question is: how would a different system do? To find out, we
use three different systems for comparison. When
``recommending'' at \emph{random}, news are simply chosen at
random from the pool of available news. When recommending by
\emph{absolute popularity}, a news is recommended according to
the number of users who approved it. When recommending by
\emph{relative popularity}, a news is recommended according to
the ratio of its number of approvals to the number of its
evaluations. In Fig.~\ref{fig:comparison}a, we compare the three
simple systems with our adaptive model for various values of the
acceptance threshold $\Delta$ (the lower the threshold, the less
demanding the users). As can be seen, our system outperforms the
others over a wide range of $\Delta$. Only when users demand
little ($\Delta\lesssim3$), recommendation by relative
popularity is able to work similarly well. However, notice that
performance of popularity-based systems is strongly influenced
by how much users differ in their tastes---this effect is shown
in Fig.~\ref{fig:comparison}b where $6$ active tastes out of
$24$ are assumed. Within the described artificial framework, one
can test also the correlation-based recommendation method by
Resnick \etal~\cite{RISBR94}. Our results show that the learning
phase of this method is longer and the resulting performance is
worse than those achieved with the adaptive model.

\begin{figure}
\centering
\includegraphics[scale=0.25]{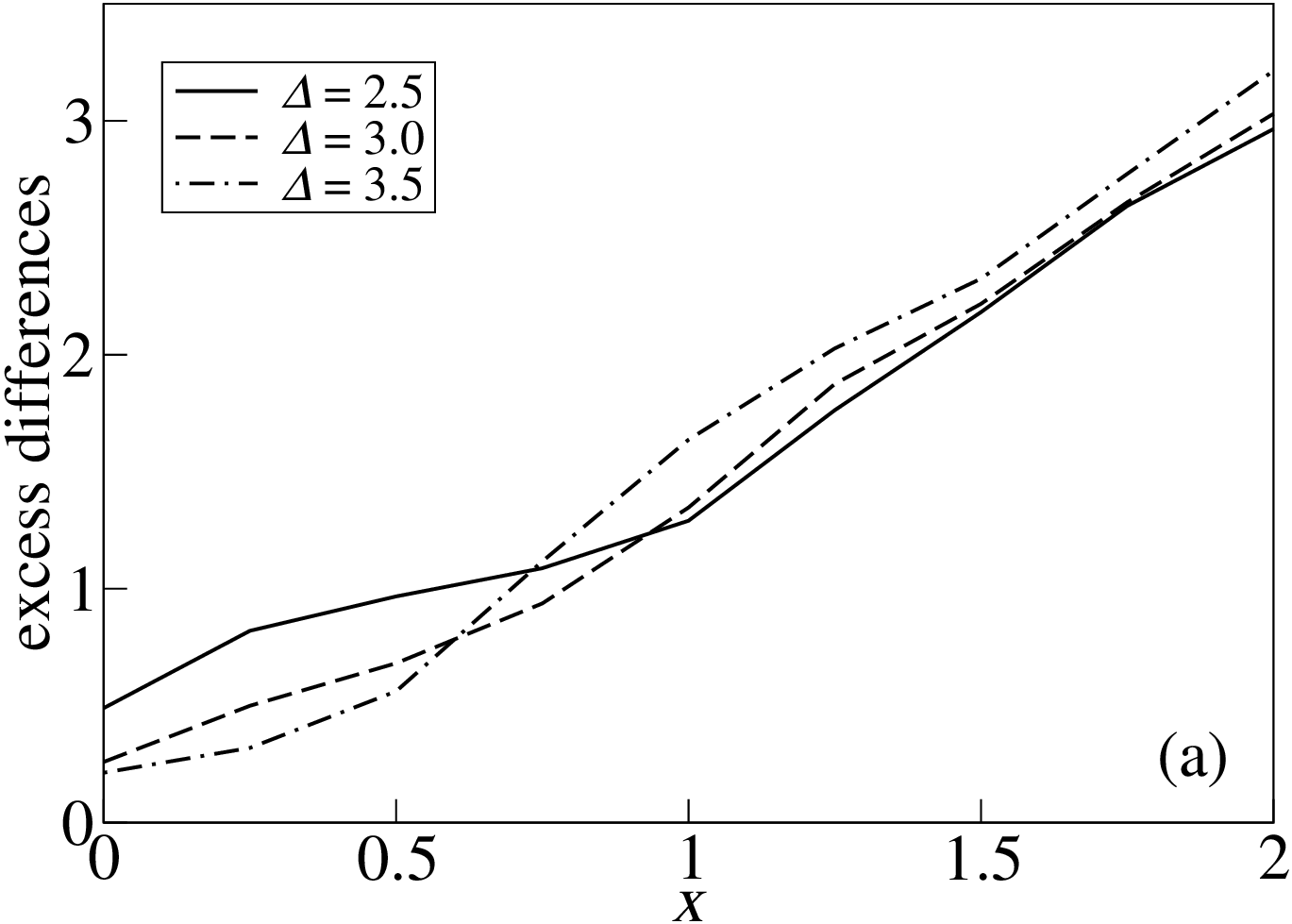}\\[8pt]
\includegraphics[scale=0.25]{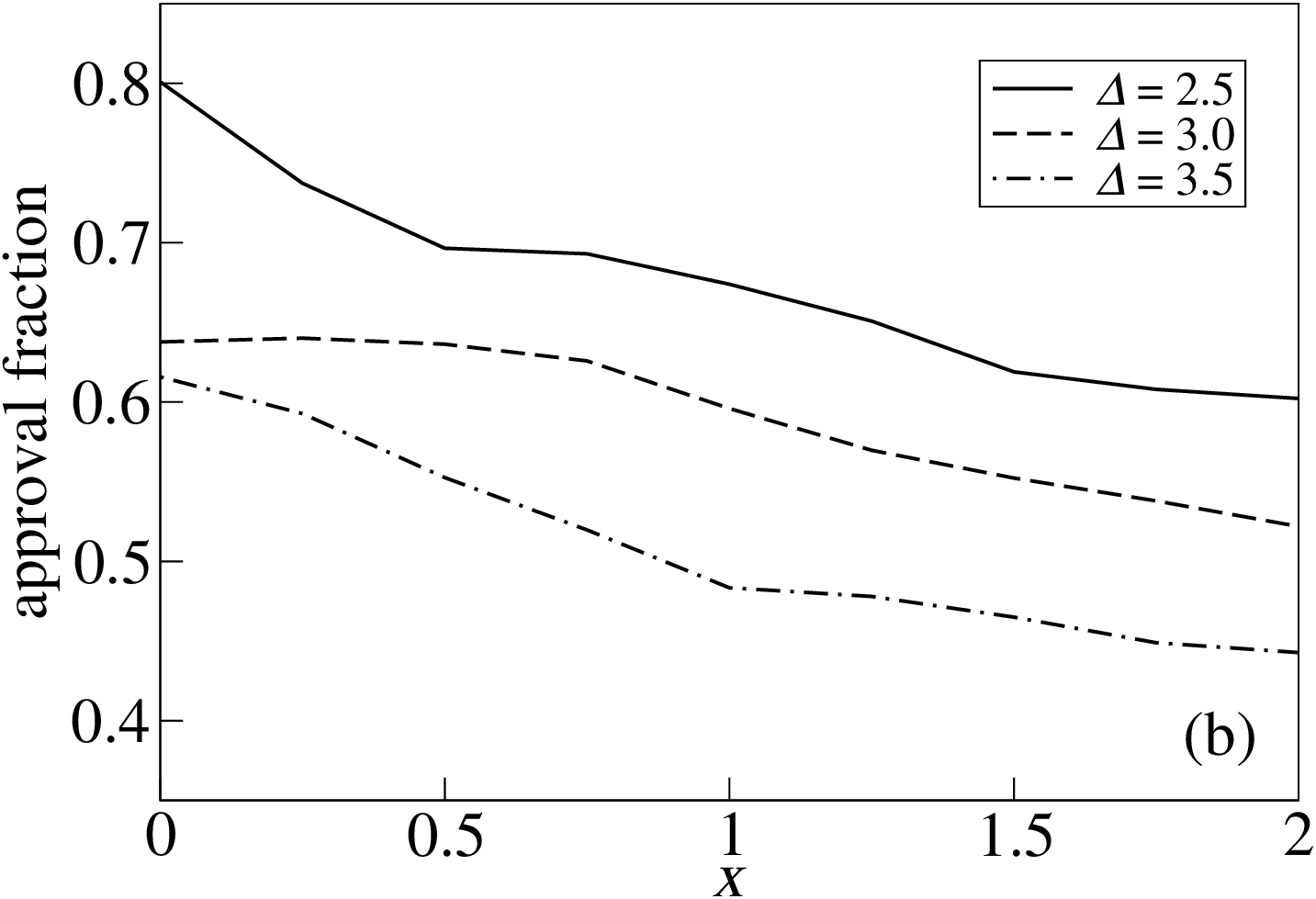}
\caption{Dependency of the system's performance on the amplitude
of users' evaluation errors $x$.}
\label{fig:noisy_evaluation}
\end{figure}

Real people are not judging machines and hence unintentional
errors are always present in their evaluations (intentional bias
is often a problem too, we discuss it later). To include these
errors in our simulations, we generalize Eq.~(\ref{overlap}) to
the form
\begin{equation}
\label{overalp-noisy}
\Omega(i,\alpha)=Q_{\alpha}(\boldsymbol{t}_i,\boldsymbol{a}_{\alpha})+xE
\end{equation}
where $E$ is a random variable drawn from the uniform
distribution $\mathcal{U}(-1,1)$ and $x>0$ is the error
amplitude. As shown in Fig.~\ref{fig:noisy_evaluation},
evaluation errors have negative influence on the system's
performance. However, while the number of excess differences
grows greatly (in Fig.~\ref{fig:noisy_evaluation}, the increase
is more than ten-fold), the approval fraction, which is a more
practical measure, is much less sensitive (in
Fig.~\ref{fig:noisy_evaluation}, the decrease is less than
$20\%$). We can conclude that the presented system is rather
robust with respect to unintentional evaluation errors.

Real users are heterogeneous not only in their tastes (as we
have studied above) but also in the frequency and pattern of
system's usage, in the threshold levels of news judgment, in the
amplitude of judgment errors, and other aspects. These effects
are easy to be studied within the presented framework. For
example, our simulations show that noisy users have less
followers than more careful users. The frequency of usage, while
very important in the initial learning phase (when heavy users
have more followers than casual users), is of little importance
when the system has approached the optimal assignment of
authorities.

\section{Discussion}
We introduced a novel news recommender model and studied its
behavior and performance in an artificial environment. We tried
to keep the model as simple as possible, yet not sacrificing its
performance to simplicity. For example, one can think of
replacing the maximization of the similarity $s_{ij}$ with
a~more sophisticated technique for the selection of authorities.
We tested a technique based on the factorization of the matrix
of users' ratings~\cite{gravity07} but despite substantially
higher computational complexity, the improvement obtained with
this method is none or little. Yet, the possibility of merging
the presented recommendation model with a~different method by
\emph{hybridization}~\cite{Burke02} remains open.

Apart from the agent-based simulations presented here, we would
like to discuss some aspects of the model's application in real
life. For any information filtering technique, its vulnerability
to malicious behavior is of crucial importance. Hence it is
important to notice that the presented system is resistant to
actions of spammers. To illustrate this, let's imagine that a
spammer introduces a new junk news to the system. Two things
happen then. First, the news is sent to a small number of the
spammer's followers (if there are some) and after it is
disapproved by them, the news is effectively removed from the
system after ``harming'' only a handful of users. Second,
spammers tend to disagree with their followers (who dislike
their spam news) and hence they loose these followers fast and
soon are left without any influence on the system at all.
Surprisingly, a similar thing would happen to a renowned news
agency which would decide to act as a user and feed the system
with the agency's news. Since agencies usually produce news
covering many different areas, most users would find a large
fraction of those news not interesting and the system would
attach them to another users with more refined preferences and
hence a higher similarity value. In other words, our model
favors ``selective sources'' of information over high-quality
non-selective sources.

In any real application of the model, there are many technical
issues which need to be addressed. For example, the initial
random assignment of authorities can be easily improved when
users are asked to provide some information about their
preferences. This information can be transformed to a
semi-optimal initial assignment which is further improved on the
basis of users' evaluations. There is also the cold start
problem: at the beginning, most users have no news recommended
(the same holds also later for fresh users). To overcome this,
one could merge the proposed spreading-based recommendation model
with simple popularity based recommendation. Further, users may
be given the possibility to choose some or all of their
authorities by themselves. While hard to model in a computer
agent-based simulation, this freedom of choice may significantly
improve users' satisfaction and their trust in the system. The
recent popularity of online social applications tells us that
regardless how sophisticated an mathematical algorithm is, users
often prefer recommendations from sources whom they know and
trust~\cite{SwSi01}. Finally, one can object that in our model,
reputation of the user who introduces a news to the system is of
zero importance. In practice it is easy to reflect this
reputation by, for example, increasing the recommendation score
of news introduced by users with a good submission record.

The ultimate test of the system's performance and viability can
be provided only by tests with real users. We are looking
forward to this stage of our research.

\acknowledgements We acknowledge stimulating discussions with
Giulio Cimini. This work was partially supported by Swiss
National Science Foundation (grant no.~200020-121848), T.Z.
acknowledges support of the National Natural Science Foundation
of China (grant nos. 60744003 and 10635040).

\end{document}